\documentclass[twocolumn,english,aps,prl,reprint,floatfix,notitlepage,nofootinbib,preprintnumbers,superscriptaddress,longbibliography]{revtex4-1}
\pdfoutput=1
\usepackage{lmodern}

\usepackage[T1]{fontenc}
\usepackage[latin9]{inputenc}
\usepackage{geometry}
\usepackage{subfigure,lmodern, amsmath,amssymb, graphicx, pifont, adjustbox, bm, xcolor}
\usepackage{amsfonts}
\usepackage{enumitem}
\usepackage{comment}
\usepackage{mathtools}
\usepackage{float}
\usepackage{slashed}
\usepackage{ragged2e}
\usepackage{array}
\usepackage{bbm}
\usepackage{balance}
 \usepackage{booktabs}
\usepackage{multirow}
\usepackage{tikz}
\usetikzlibrary{decorations.markings,arrows.meta,calc}

\usepackage{nameref}

\usepackage{hhline}

\makeatletter\g@addto@macro\bfseries{\boldmath}\makeatother

\makeatletter\newcommand{\labeltext}[2]{%
  \def\@currentlabel{#1}%
  \label{#2}%
}
\makeatother

\newcommand{\GammaTwo}{\Gamma_{\!2}}
\newcommand{\GammaFour}{\Gamma_{\!4}}

\usepackage{stackengine}
\usepackage{esint}
\usepackage[unicode=true,pdfusetitle,
 bookmarks=true,bookmarksnumbered=false,bookmarksopen=false,
 breaklinks=false,pdfborder={0 0 1},backref=false,colorlinks=true]
 {hyperref}
\hypersetup{
 pdfauthor={Francesco Calisto,Clifford Cheung},
 citecolor=blue,linkcolor=black,urlcolor=blue}

\newcommand{\appendixref}[1]{\hyperref[#1]{appendix~\ref{#1}}}
\def\equationautorefname~#1\null{eq.\,(#1)\null}
\usepackage{breakurl}
\usepackage[hang,flushmargin]{footmisc} 
\allowdisplaybreaks
\makeatletter

\usepackage{etoolbox}
\apptocmd{\thebibliography}{\justifying\setlength{\leftskip}{7.4mm}}{}{} 
 
 \usepackage{relsize}
\usepackage{babel}

\makeatletter
\def\simgt{\mathrel{\lower2.5pt\vbox{\lineskip=0pt\baselineskip=0pt
           \hbox{$>$}\hbox{$\sim$}}}}
\def\simlt{\mathrel{\lower2.5pt\vbox{\lineskip=0pt\baselineskip=0pt
           \hbox{$<$}\hbox{$\sim$}}}}
\makeatother

\usepackage{changepage}

\newcommand{\be}{\begin{equation}}
\newcommand{\ee}{\end{equation}}
\newcommand{\bea}{\begin{eqnarray}}
\newcommand{\eea}{\end{eqnarray}}
\newcommand{\Fig}[1]{Fig.~\ref{#1}}

\newcommand{\Eq}[1]{Eq.\,(\ref{#1})}
\newcommand{\Eqs}[2]{Eqs.\,(\ref{#1}) and (\ref{#2})}

\newcommand{\eq}[2]{\be\begin{aligned}#1 \label{#2}\end{aligned}\ee}

\newcommand{\mysec}[1]{\noindent {\bf #1.}---}

\newcolumntype{P}[1]{>{\centering\arraybackslash}p{#1}}

\usepackage{fix-cm}

\definecolor{dartmouthgreen}{rgb}{0.05, 0.5, 0.06}

\begin{document}

\preprint{CALT-TH 2026-024}

\title{An Ultraviolet Finite Theory of Scalars}

\author{Francesco Calisto}
\affiliation{Walter Burke Institute for Theoretical Physics and
Leinweber Forum for Theoretical Physics, California Institute of Technology, Pasadena, CA 91125, USA}

\author{Clifford Cheung}
\affiliation{Walter Burke Institute for Theoretical Physics and
Leinweber Forum for Theoretical Physics, California Institute of Technology, Pasadena, CA 91125, USA}

\begin{abstract}
\noindent We construct a theory of scalars that is free of short-distance infinities
to all orders in perturbation theory.  Loop divergences are neutralized by momentum-dependent
interactions that are ghost free and polynomially bounded.  The finite counterparts
of the usual one-loop scalar self-energy and beta function are straightforwardly
computed.  In a variant of this model, the one-loop mass renormalization is
zero due to an inversion that swaps the ultraviolet and the
infrared.
\end{abstract}

\maketitle

\mysec{Introduction}High-energy physics is, at its core, the study of ever shorter distances.
Yet its primary instrument---effective
field theory---is beset by infinities.  Explicit calculations generate
ultraviolet divergences that must be regularized and renormalized in order to
make physical predictions.

Of course, there are exceptional theories that are completely finite to
all orders in perturbation theory.  In four dimensions, all such
examples rely on additional structures like conformal symmetry,
supersymmetry, or both.  Absent these protections, a more fraught strategy is
to introduce nonlocality directly into the dynamics.

The archetype of this approach is the Lee-Wick mechanism, which forcibly tames
the ultraviolet using higher-derivative kinetic terms
\cite{LeeWick1969,LeeWick1970,Cutkosky1969,LeeWick1971,GrinsteinOConnellWise2008}.
These corrections induce a modified propagator,
\eq{
\frac{1}{\Box(1+\Box/m^2)}
=
\frac{1}{\Box}-\frac{1}{\Box+m^2},
}{eq:LW-propagator}
whose sole purpose is to render divergent loop corrections finite.
Similar approaches have been applied to gravity
\cite{Stelle1977,Bach1921,Riegert1984,SalvioStrumia2014}.
In a related tactic
\cite{Moffat1990,KleppeWoodard1992,Tomboulis1997,BiswasGerwickKoivistoMazumdar2012},
interactions are dressed by momentum-dependent form
factors,
\eq{
\exp(-\Box/m^2).
}{eq:exponential-vertex}
While these mechanisms soften high-energy behavior,
they do so at an immense cost: unitarity and causality.
\Eq{eq:LW-propagator} exhibits new ghostly degrees of freedom
\cite{LeeWick1969,LeeWick1970,Stelle1977,Bach1921,Riegert1984}, while
\Eq{eq:exponential-vertex} drastically violates the polynomial
boundedness required of causal theories
\cite{Froissart1961,Martin1963,AdamsEtAl2006,CamanhoEdelsteinMaldacenaZhiboedov2016}.

In this paper we construct a four-dimensional theory of scalars
that is finite to all orders in perturbation theory.  Ultraviolet divergences
are completely absent.  By design, this theory is exorcized of ghosts and
exhibits polynomially bounded interactions.
The action for this theory, $S=S_0+S_{\rm int}$, has a canonical
kinetic term $S_0=-\frac12\int d^4x\,\phi\Box\phi$
together with a quartic interaction,
\eq{
S_{\rm int}
&=\frac1{4!}\int\limits_{p_1\cdots p_4}
G(p_1,\ldots,p_4)\,\phi(p_1)\cdots\phi(p_4),
}{eq:action}
where we have defined the momentum-space integral,
\eq{
\int\limits_{\mathclap{p_1\cdots p_4}}
&=
\prod_{i=1}^4\int\!d^4p_i\,
\delta^{(4)}\!\left(\sum_{i=1}^4 p_i\right).
}{eq:int-measure}
Here the interactions are controlled by a momentum-dependent form
factor\footnote{Here we have defined the off-shell Mandelstam invariants,
$s=(p_1+p_2)^2$, $t=(p_2+p_3)^2$, and $u=(p_3+p_1)^2$, where $s+t+u=0$ for
on-shell kinematic configurations.},
\eq{
G(p_1,p_2,p_3,p_4)=A(s,t,u),
}{eq:G-A-def}
interpreted throughout as the off-shell continuation of a physical two-to-two
scattering amplitude.

Conveniently, there is a longstanding endeavor within the modern
amplitudes program, known colloquially as ``positivity''
\cite{AdamsEtAl2006,BellazziniMiroRattazziRiembauRiva2020,ArkaniHamedHuangHuang2020,CaronHuotMazacRastelliSimmonsDuffin2021}, whose entire mission
statement is to fully characterize the space of functions $A(s,t,u)$ consistent
with unitarity and causality.  Taking direct inspiration from those recent
developments, we devise a new class of scalar theories with many curious
features.

First, we show how to trivially construct
momentum-dependent interactions that are ghost free and polynomially bounded.
Second, by interpreting the kinematic singularities of the interactions as
heavy intermediate states, we learn that each one carries {\it infinite spin}.  While
reminiscent structures appear in string theory, in the present context each massive
resonance exhibits its very own infinite-spin tower.
Furthermore, all of our models require a negative quartic potential.  Strictly
speaking, this is allowed---and perhaps even favored by current
measurements of the Higgs boson and top quark masses
\cite{DegrassiEtAl2012,ButtazzoEtAl2013,AndreassenFrostSchwartz2014}---but it indicates a classically unstable minimum.

These idiosyncrasies strongly suggest that these theories are
unlikely to describe the world we live in.  But they offer one
undeniable virtue: they are provably free of ultraviolet divergences at all
loop orders.  By explicit calculation, we verify that all one-loop corrections are
divergence-free, and in doing so reproduce the finite analogs of the scalar
quadratic divergence and logarithmic running.  We also describe a theory in
which the one-loop mass correction is exactly
zero.  This arises because the loop integrand is a total derivative and
exhibits a mysterious inversion that cancels the ultraviolet and
infrared contributions to the loop.

\medskip
\mysec{Ultraviolet Finiteness}
Typically, divergences first appear in the
one-loop self-energy diagram,
\eq{
-i\GammaTwo(p^2)&=
\vcenter{\hbox{%
\begin{tikzpicture}[
line width=0.95pt,
scale=1.15,
momentum/.style={postaction={decorate},
decoration={markings, mark=at position #1 with {\arrow{Stealth[length=3pt,width=4pt]}}}},
loopmomentum/.style={postaction={decorate},
decoration={markings, mark=at position #1 with {\arrow[rotate=-10]{Stealth[length=3pt,width=4pt]}}}}
]
\path[use as bounding box] (-1.35,-0.98) rectangle (1.35,0.98);
\begin{scope}[yshift=-0.42cm]
\coordinate (V) at (0,0);
\coordinate (L) at (-1.05,0);
\coordinate (R) at (1.05,0);
\draw[momentum=0.55] (L) -- (V);
\draw[momentum=0.55] (V) -- (R);
\draw[loopmomentum=0.55] (V) .. controls (1.12,1.12) and (-1.12,1.12) .. (V);
\draw[fill=lightgray] (V) circle (0.105);
\node at (0,1.18) {$\ell$};
\node at (-0.73,0.28) {$p$};
\node at (0.73,0.28) {$p$};
\end{scope}
\end{tikzpicture}
}}.
}{eq:self-energy-diagram}
Here the quartic vertex is evaluated at $s=(p+\ell)^2$, $t=0$, and
$u=(p-\ell)^2$, yielding an integral over the amplitude,
\eq{
-i\GammaTwo(p^2)
&=
-\frac12\int\!\frac{d^4\ell}{(2\pi)^4}\,
\frac{1}{\ell^2}\,
A\!\left((p+\ell)^2,0,(p-\ell)^2\right).
}{eq:Gamma2-p2}
Analytically continuing to Euclidean space,
\eq{
\ell^2\to -q^2
\quad\text{and}\quad
\int\!\frac{d^4\ell}{(2\pi)^4}\frac{1}{\ell^2}
\to -\frac{i}{16\pi^2}\int_0^\infty dq^2,
}{eq:ell-q-measure}
we find that the self-energy takes the form
\eq{
\GammaTwo(0)=\frac{\lambda\Lambda^2}{32\pi^2},
}{eq:Lambda-def}
where we have defined the low-energy quartic coupling,
\eq{
\lambda=-A(0,0,0),
}{eq:lambda-def}
together with the effective ultraviolet cutoff
\eq{
\Lambda^2
=
\int_0^\infty dq^2\,
\frac{A(-q^2,0,-q^2)}{A(0,0,0)}.
}{eq:Lambda-master}
Thus, the one-loop self-energy diagram is finite provided the
amplitude falls off appropriately, so
\eq{
\lim_{s\to\infty}sA(s,0,s)=0.
}{eq:UV-euclidean-falloff}
All of our models will satisfy this condition\footnote{The very definition of ultraviolet completion is a high-energy continuation that softens the dynamics. Without this, short-distance behavior is unchanged, calling into question what has actually been achieved.  For a low-energy quartic potential, softening requires an amplitude that falls off faster than the constant quartic coupling $\lambda$, corresponding to a vanishing Regge limit.  Using a standard unsubtracted dispersion relation, it is then straightforward to show that unitarity implies $\lambda\leq 0$ in generality.}.

What are the sufficient conditions for finiteness at all orders in perturbation
theory?  First, recall
that the quartic coupling in canonical $\phi^4$ theory is precisely marginal.
Consequently, any softening of the quartic interaction at
high energies will soften the degree of divergence, rendering the theory finite.

For generic off-shell external kinematics, this translates to the requirement
that
\eq{
\lim_{s,t,u\to\infty}A(s,t,u)=0.
}{eq:UV-limit}
Crucially, the interaction vertex must also be softened at more
special kinematics.  \Eq{eq:UV-euclidean-falloff} ensures that this is the case
when one Mandelstam invariant is zero.  However, when two are zero, we will
typically find that
\eq{
\lim_{s\to\infty}A(s,0,0)\sim \text{constant}.
}{eq:UV-exceptional-channel}
Because this scales precisely like the quartic coupling of
$\phi^4$ theory, it naively
reintroduces ultraviolet divergences.  

Fortunately, finiteness is nevertheless robust.  To see why, consider any connected hard
subdiagram,
\eq{
\begin{tikzpicture}[
baseline=-0.55cm,
scale=1.05,
line width=0.95pt,
hard/.style={line width=0.95pt},
softleg/.style={line width=0.95pt,draw=gray!70},
leg/.style={softleg},
soft/.style={softleg,dotted},
boundary/.style={draw=gray!22,line width=0.8pt}
]
\coordinate (A) at (-0.70,0.55);
\coordinate (B) at (0.70,0.55);
\coordinate (C) at (0.70,-0.55);
\coordinate (D) at (-0.70,-0.55);
\coordinate (Ea) at (-1.28,0.92);
\coordinate (Eb) at (0.18,1.57);
\coordinate (Ec) at (1.37,0.78);
\coordinate (Ed) at (1.28,-0.92);
\coordinate (Ee) at (-0.18,-1.57);
\coordinate (Ef) at (-1.37,-0.78);
\coordinate (Oa) at ($(A)!1.55!(Ea)$);
\coordinate (Ob) at ($(B)!1.50!(Eb)$);
\coordinate (Oc) at ($(B)!1.55!(Ec)$);
\coordinate (Od) at ($(C)!1.55!(Ed)$);
\coordinate (Oe) at ($(D)!1.50!(Ee)$);
\coordinate (Of) at ($(D)!1.55!(Ef)$);
\draw[boundary] (0,0) circle (1.48);
\draw[hard] (A) -- (B);
\draw[hard] (B) -- (C);
\draw[hard] (C) -- (D);
\draw[hard] (D) -- (A);
\draw[hard] (A) -- (C);
\draw[leg] (A) -- (Ea);
\draw[leg] (B) -- (Eb);
\draw[leg] (B) -- (Ec);
\draw[leg] (C) -- (Ed);
\draw[leg] (D) -- (Ee);
\draw[leg] (D) -- (Ef);
\draw[soft] (Ea) -- (Oa);
\draw[soft] (Eb) -- (Ob);
\draw[soft] (Ec) -- (Oc);
\draw[soft] (Ed) -- (Od);
\draw[soft] (Ee) -- (Oe);
\draw[soft] (Ef) -- (Of);
\draw[fill=lightgray] (A) circle (0.115);
\draw[fill=lightgray] (B) circle (0.115);
\draw[fill=lightgray] (C) circle (0.115);
\draw[fill=lightgray] (D) circle (0.115);
\end{tikzpicture}
}{eq:hard-subgraph-diagram}
whose internal lines all carry hard momenta by definition.  The boundary of this subdiagram is delineated by the transition to soft external or loop momenta.  
  Any boundary vertex carries two
or three hard incoming momenta, so it is evaluated in the kinematic regime of
\Eq{eq:UV-euclidean-falloff} rather than 
\Eq{eq:UV-exceptional-channel}.  Assuming the falloff in
\Eq{eq:UV-euclidean-falloff}, every hard subdiagram has at least one vertex
softened relative to $\phi^4$ theory\footnote{Here we have assumed that the
diagram in question has at least one external leg.  For vacuum diagrams, this
argument fails, so the cosmological constant is ultraviolet divergent.}, so the hard subdiagram is
no longer marginal but ultraviolet finite.

The above logic can be made quantitative by standard methods \cite{Weinberg1960}. For a four-dimensional theory with quartic interactions, any $L$-loop subdiagram with $I$ propagators and $E$ external legs has a superficial degree of divergence $\Delta=4L-2I-S=4-E-S$, where $S$ is the total degree of momentum suppression relative to $\phi^4$ theory.  Assuming \Eqs{eq:UV-euclidean-falloff}{eq:UV-limit}, this implies that $S> 2$, so $\Delta< 2 -E$.  Hence, any diagram with at least two external legs will be ultraviolet finite. For a detailed accounting of divergences, including subintegration regions, see App.~A.

In summary, for any theory in which \Eqs{eq:UV-euclidean-falloff}{eq:UV-limit}
are satisfied, there are no ultraviolet divergences at any perturbative loop
order.
We emphasize that these conditions are required to hold at complex kinematics.
As a result, these conditions exclude momentum-dependent vertices constructed directly from string
amplitudes, whose hard-scattering behavior is
$\log A \sim -(s\log s+t\log t+u\log u)$ \cite{CaronHuotKomargodskiSeverZhiboedov2017}.
Such amplitudes diverge exponentially in some complex kinematic direction.

\medskip
\mysec{Vertex Ansatz}The quartic interaction of this scalar theory is dictated by a scattering
amplitude that we define by the following general ansatz,
\eq{
A(s,t,u)=\frac{N(s,t,u)}{D(s,t,u)}.
}{eq:general-rational}
For simplicity, we assume throughout that \Eq{eq:general-rational}
is a rational function, though this is not required by consistency.  We further
assume a tree-level structure where the denominator is the product of poles,
\eq{
D(s,t,u)
=
\prod_n(m_n^2-s)(m_n^2-t)(m_n^2-u).
}{eq:numerator-general}
Like in Lee-Wick theories, we can interpret these poles as intrinsic kinematic features of the form factor interaction
or as bona fide exchanged states.  In the latter viewpoint, the denominator encodes the
crossing-symmetric exchange of a tower of heavy states labeled by an index
$n$, with masses $m_n^2$.
Without loss of generality, the numerator is a crossing-symmetric
function,
\eq{
N(s,t,u)
=
N(\sigma_2,\sigma_3),
}{eq:sigma-def}
where
$\sigma_2=\frac12(s^2+t^2+u^2)$ and $\sigma_3=stu$.  We assume throughout
that the numerator is a polynomial.

From \Eqs{eq:lambda-def}{eq:Lambda-master}, we obtain the
quartic,
\eq{
\lambda
=
-\frac{N(0,0)}{\displaystyle\prod_n m_n^6},
}{eq:lambda-general}
while the effective cutoff is
\eq{
\Lambda^2
=
\int_0^\infty dq^2\,
\frac{
N(q^4,0)/N(0,0)
}{
\displaystyle
\prod_n\left(1+\frac{q^2}{m_n^2}\right)^2
},
}{eq:Lambda-general}
which will be useful when we consider concrete models.

As we saw earlier, \Eqs{eq:UV-euclidean-falloff}{eq:UV-limit} are required for the theory to be finite at all perturbative orders.  These
constrain the relative degrees of the numerator and denominator in
\Eq{eq:general-rational},
\eq{
\deg N(s,t,u)
&<
\deg D(s,t,u),
\\
\deg N(s,0,s)+1
&<
\deg D(s,0,s),
}{eq:degree-finiteness}
which is not a particularly stringent constraint.

Note that our ansatz carries no dependence on the external momenta squared,
\eq{
p_i^2\overset{\raisebox{.35ex}{$\scriptstyle\rm on\mbox{-}shell$}}{\longrightarrow}0,
}{eq:on-shell-limit}
which vanish on-shell but can in principle enter into
the off-shell continuation of the amplitude.  For example, we could choose an
off-shell continuation of the amplitude so that such terms appear in
propagator denominators,
\eq{
\frac{1}{m_n^2-s-\sum_i c_i p_i^2}
\overset{\raisebox{.35ex}{$\scriptstyle\rm on\mbox{-}shell$}}{\longrightarrow}
\frac{1}{m_n^2-s},
}{eq:on-shell-pole}
thus softening the high-energy behavior in off-shell loops even more.
However, it is
worth emphasizing that such denominator contributions will generically spoil
higher-point factorization, since they introduce spurious poles in kinematic
regions that can never arise from the exchange of physical states.  In any
case, we will not consider factors of $p_i^2$ in the form factor any further
here.

\medskip
\mysec{Simple Example}The minimal scattering amplitude that softens the ultraviolet while remaining
ghost free and polynomially bounded is
\eq{
A(s,t,u)
=
-\frac{\lambda m^6}{(m^2-s)(m^2-t)(m^2-u)},
}{eq:one-mass-A}
where $\lambda<0$.  This peculiar object was first discovered in the study of scattering
amplitudes \cite{CaronHuotVanDuong2021}, and to date there is a rich family of generalizations \cite{HuangRemmen}.  As we will see, it is by design ghost free\footnote{While these theories are ghost free, they also require $\lambda<0$, which of course
has its own classical instability.  But in perturbation theory this pathology
is invisible.}
and polynomially bounded.  It is worth noting that the entire raison d'\^etre of
the positivity program in amplitudes is to derive form factors satisfying these
conditions.
Since \Eq{eq:one-mass-A} satisfies
\Eqs{eq:UV-euclidean-falloff}{eq:UV-limit}, it defines an ultraviolet finite
theory.

\medskip
\noindent {\it Radiative Corrections.} From \Eq{eq:lambda-def}, we see that
\Eq{eq:one-mass-A} defines an ultraviolet completion of $\phi^4$ theory that is
finite to all perturbative orders.  For this reason, it will be illuminating to compute the
one-loop corrections to the self-energy and the quartic interaction.

From \Eq{eq:Gamma2-p2}, the one-loop self-energy diagram is
\eq{
-i\GammaTwo(p^2)
&=
\frac{\lambda m^4}{2}
\int\!\frac{d^4\ell}{(2\pi)^4}\,
\frac{1}{
\ell^2[(\ell+p)^2-m^2]}
\\
&\quad\times
\frac{1}{[(\ell-p)^2-m^2]}
\\
&=
-i\,\frac{\lambda m^2}{32\pi^2}
\int_0^1\!d\alpha\int_0^{1-\alpha}\!d\beta\,
\\
&\quad\times
\frac{1}{
\alpha+\beta
-(p^2/m^2)[\alpha+\beta-(\alpha-\beta)^2]
}.
}{eq:one-mass-Gamma2-finite-p-integral}
After a slight change of variables in the Feynman parameters\footnote{%
The integral simplifies drastically after going to new variables,
$u=\alpha+\beta$ and $v=\alpha-\beta$.},
we obtain the finite expression
\eq{
\GammaTwo(p^2)
=
\frac{\lambda m^2}{32\pi^2}\,I_2\!\left(\frac{p^2}{m^2}\right),
}{eq:one-mass-Gamma2-finite-p-closed}
where we have defined
\eq{
I_2(x)
=
\frac{\log(1-x)}{x}
+\frac{2}{\sqrt{x(1-x)}}\,
\tan^{-1}\!\sqrt{\frac{x}{1-x}}.
}{eq:one-mass-finite-p-F}
The relevant limits of $I_2(x)$ are
\eq{
I_2(x)
=
\begin{dcases}
1+\frac{5}{6}x+\mathcal O(x^2),
& x\to0,
\\
-\frac{\log4}{x}
+\mathcal O(x^{-2}),
& x\to-\infty
\end{dcases}.
}{eq:one-mass-F-limits}
Thus, at low energies, the self-energy contributes a finite mass
correction,
\eq{
\GammaTwo(0)=\frac{\lambda m^2}{32\pi^2},
}{eq:one-mass-Gamma2-final}
while at high energies it asymptotes to zero.
The finiteness of this one-loop correction follows from
\eq{
\Lambda^2
=
\int_0^\infty dq^2\,
\frac{1}{\left(1+q^2/m^2\right)^2}
=
m^2,
}{eq:one-mass-Lambda}
which says that the divergence is cut off by the mass of the
exchanged state.

Next, let us compute the one-loop correction to the quartic
interaction,
\eq{
-i\GammaFour(s,t,u)
&=\quad
\begin{tikzpicture}[
baseline=-0.10cm,
line width=0.95pt,
scale=1.15,
momentum/.style={postaction={decorate},
decoration={markings, mark=at position #1 with {\arrow{Stealth[length=3pt,width=4pt]}}}},
loopmomentum/.style={postaction={decorate},
decoration={markings, mark=at position #1 with {\arrow{Stealth[length=3pt,width=4pt]}}}}
]
\coordinate (L) at (-0.50,0);
\coordinate (R) at (0.50,0);
\coordinate (Pone) at (-1.30,0.52);
\coordinate (Ptwo) at (-1.30,-0.52);
\coordinate (Pthree) at (1.30,0.52);
\coordinate (Pfour) at (1.30,-0.52);
\draw[momentum=0.55] (Pone) -- (L);
\draw[momentum=0.55] (Ptwo) -- (L);
\draw[momentum=0.55] (Pthree) -- (R);
\draw[momentum=0.55] (Pfour) -- (R);
\draw[loopmomentum=0.55] (L) .. controls (-0.50,0.67) and (0.50,0.67) .. (R);
\draw[loopmomentum=0.55] (R) .. controls (0.50,-0.67) and (-0.50,-0.67) .. (L);
\draw[fill=lightgray] (L) circle (0.105);
\draw[fill=lightgray] (R) circle (0.105);
\node at (0,-0.78) {$\ell$};
\node at (-1.12,0.78) {$p_1$};
\node at (-1.12,-0.78) {$p_2$};
\node at (1.12,0.78) {$p_3$};
\node at (1.12,-0.78) {$p_4$};
\end{tikzpicture}
+\cdots,
\\
&=
-\frac{i}{2}\left[
F(s,t,u)+F(t,u,s)+F(u,s,t)
\right],
}{eq:Gamma4-diagram}
where the $s$-channel contribution is
\eq{
-iF(s,t,u)
&=
\frac{\lambda^2m^{12}}{(m^2-s)^2}
\int\!\frac{d^4\ell}{(2\pi)^4}
\frac{1}{\ell^2(\ell+p_1+p_2)^2}
\\
&\quad\times
\frac{1}{
[(\ell+p_1)^2-m^2][(\ell+p_2)^2-m^2]}
\\
&\quad\times
\frac{1}{
[(\ell-p_3)^2-m^2][(\ell-p_4)^2-m^2]}.
}{eq:one-mass-four-point-setup}
While this integral is difficult to evaluate in generality, it is tractable at the
symmetric point, $
s=t=u=-\mu^2$,
corresponding to $p_i^2=-\frac{3\mu^2}{4}$ and $
p_i p_j=\frac{\mu^2}{4}$.  The sum over channels then collapses to
$\GammaFour(-\mu^2)=\frac32\,F(-\mu^2)$.

At energies far below the gap, we find that
\eq{
-iF(-\mu^2)
&\overset{\raisebox{.35ex}{$\scriptstyle\mu^2\ll m^2$}}{=}
\lambda^2m^8\!\int\!\frac{d^4\ell}{(2\pi)^4}
\\[-1mm]
&\quad\times
\frac{1}{\ell^2(\ell+p_1+p_2)^2(\ell^2-m^2)^4}
\\[-1mm]
&\hspace{-1.2cm}=
\frac{i\lambda^2}{16\pi^2}\!\int_0^1\!d\alpha\int_0^{1-\alpha}\!d\beta\,
\beta^3\,[\beta+(\mu^2/m^2)\alpha(1-\alpha)]^{-4},
}{eq:one-mass-four-point-small-mu-lorentzian}
which, after a change of variables\footnote{Specifically, we
transform to $u=\beta/(1-\alpha)$ and $v=(\mu^2/m^2)\alpha$.}, evaluates to
\eq{
\GammaFour(-\mu^2)
&\overset{\raisebox{.35ex}{$\scriptstyle\mu^2\ll m^2$}}{=}
\frac{3\lambda^2}{32\pi^2}
\left[
\log\frac{\mu^2}{m^2}
+\frac56
\right].
}{eq:one-mass-four-point-small-mu-final}
The above formula reproduces the renormalization of $\phi^4$ theory, which is
well approximated when $\mu^2\ll m^2$.  The log
correctly encodes the expected positive one-loop beta function coefficient.  If
extrapolated indefinitely, such running
naively terminates at a Landau pole, signaling a breakdown of
perturbation theory.  In the present case this conclusion is avoided twice over.  First, as
we will see below, positivity requires $\lambda<0$, so the same formal running
drives the coupling toward zero in the ultraviolet \cite{phi4AF}.  Second, \Eq{eq:one-mass-four-point-small-mu-final} is only the
low-energy approximation and ceases to apply near the gap.

The crossover happens when $\mu^2 \sim m^2$.  At high energies, dimensional analysis
yields
\eq{
F(-\mu^2)
&\overset{\raisebox{.35ex}{$\scriptstyle\mu^2\gg m^2$}}{=}
\lambda^2\frac{m^{12}}{\mu^4}
\int\!\frac{d^4\ell}{(2\pi)^4\ell^2(\ell+p_1+p_2)^2}
\\[-1mm]
&\quad\times
\frac{1}{(\ell+p_1)^2(\ell+p_2)^2(\ell-p_3)^2(\ell-p_4)^2}
\\[-1mm]
&\sim
\lambda^2\left(\frac{m^2}{\mu^2}\right)^6.
}{eq:one-mass-four-point-large-mu-lorentzian}
From this we obtain the correction,
\eq{
\GammaFour(-\mu^2)
&\overset{\raisebox{.35ex}{$\scriptstyle\mu^2\gg m^2$}}{\sim}
\lambda^2\left(\frac{m^2}{\mu^2}\right)^6,
}{eq:one-mass-four-point-large-mu-final}
which says that the one-loop correction to the quartic interaction
actually decouples at sufficiently high energies, consistent with the ultraviolet
finiteness of the theory.

\medskip
\noindent {\it Unitarity Bounds.} It is easy to verify that this
theory is free of ghosts.  The residue at the massive pole is
\eq{
R(t)
&=
\lim_{s\to m^2}(m^2-s)A(s,t,-s-t)\\
&=
-\frac{\lambda m^6}{(m^2-t)(2m^2+t)}
=
-\frac{4\lambda m^2}{9-x^2},
}{eq:one-mass-residue}
where we have defined the angular variable
$\cos\theta=x=1+2t/m^2$.  Since \Eq{eq:one-mass-residue} is not a
polynomial in $x$, the state at $s=m^2$ comprises an infinite
tower of spins.

Unbounded spin within a narrow spectral band is odd, but it
has not yet been proven inconsistent.  On the one hand, it appears in tension
with covariant entropy bounds \cite{CEB}, which limit the number of degrees of
freedom that can couple to gravity.  On the other hand, there is no obvious
obstruction to deriving an energy-momentum tensor for these massive infinite-spin
states, or applying the graviton soft theorems \cite{Weinberg1965}.  Furthermore, infinite-spin
towers occur in the natural world, for instance in atomic spectra, as well as
in hypothetical but theoretically consistent string constructions
\cite{MaldacenaRemmen}.

The residue encodes the partial wave coefficients,
\eq{
a_\ell
&=
\frac{2\ell+1}{2}
\int_{-1}^{1}dx\,R(x)P_\ell(x)=
-\frac{4\lambda m^2(2\ell+1)}{3}Q_\ell(3),
}{eq:one-mass-partial-wave}
for even $\ell$ and otherwise vanishing.  Partial wave unitarity, which ensures the absence of ghosts, implies that $\lambda<0$, which ironically renders the vacuum nonperturbatively unstable.

The one-loop self-energy is also unitary. Given \Eq{eq:self-energy-diagram}, the optical theorem demands $\text{Im}(-\Gamma_2) \geq 0$. Taking the discontinuity of \Eq{eq:one-mass-Gamma2-finite-p-closed} for $x>1$, we get
\eq{
\text{Im} \left[ -\Gamma_2(x) \right] =-\frac{\lambda m^2}{32\pi} \left( \frac{1}{\sqrt{x(x-1)}}-\frac{1}{x} \right),
}{eq:one-mass-Gamma2-imaginary}
confirming again that $\lambda<0$.

\medskip
\mysec{Cancellations and Hierarchies}In our simple example from \Eq{eq:one-mass-Gamma2-final}, we arrived at the
expected outcome that the magnitude of the scalar mass renormalization is set
entirely by the mass of the heavy resonance.  No tuning of parameters can eliminate that
contribution.  Here we consider a next-to-minimal model in which the one-loop self-energy is not only
finite but also zero.  The corresponding amplitude is
\eq{
A(s,t,u)
=
\frac{
-\lambda c^3m^{12}+\tau m^8\sigma_2
}{
\displaystyle
\prod_{\rho=s,t,u}(m^2-\rho)(cm^2-\rho)
}.
}{eq:two-mass-A-tau}
Defining $z=q^2/m^2$, we obtain
\eq{
\GammaTwo(0)
&=
-\frac{m^2}{32\pi^2 c}
\int_0^\infty dz\,
\frac{-\lambda c^3+\tau z^2}{(1+z)^2(c+z)^2}
\\
&=
\frac{m^2}{32\pi^2}
(\lambda c^2-\tau)
\frac{c^2-1-2c\log c}{c(c-1)^3}.
}{eq:two-mass-Gamma2-tau}
For distinct poles, the second factor is nonzero because $c\ne1$, so the right-hand side vanishes only if
\eq{
\tau=\lambda c^2.
}{eq:tau-fixed}
For this choice the resulting amplitude is then
\eq{
A(s,t,u)
=
\frac{
-\lambda c^2m^8(cm^4-\sigma_2)
}{
\displaystyle
\prod_{\rho=s,t,u}(m^2-\rho)(cm^2-\rho)
},
}{eq:two-mass-A-final}
which we assume hereafter.  As shown in App.~B, partial wave unitarity holds if $\lambda\leq 0$ and $c\ge 3$.

At this point \Eq{eq:tau-fixed} is nothing more than a fine-tuning of the
ultraviolet parameters.  There is no mechanism for the elimination of the
one-loop mass correction.  However, it is somewhat interesting that the
structure of the one-loop self-energy exhibits several
rather extraordinary features to accommodate this zero.  In particular, the
vanishing of the effective cutoff,
\eq{
\Lambda^2
=
m^2\int_0^\infty dz\,f(z)
=
0,
}{eq:two-mass-Lambda}
follows because the loop integrand is a total derivative,
\eq{
f(z)
=
c\frac{d}{dz}
\left[
\frac{z}{(1+z)(c+z)}
\right].
}{eq:two-mass-line-final}
A similar phenomenon was observed in Ref.~\cite{ArkaniHamedHarigaya2021}.
Furthermore, the integrand flips sign under the inversion,
\eq{
f\!\left(\frac{c}{z}\right)
=
-\frac{z^2}{c}f(z),
}{eq:f-inversion}
which precisely cancels the contributions from the ultraviolet and
infrared regions of the loop.  Whether this signifies a deep principle,
persisting beyond one loop, we
leave to future inquiry.
 In particular, a rational numerator ansatz allows higher powers of crossing-symmetric polynomials, which may provide enough freedom to cancel higher-loop self-energy corrections. It would be interesting to study these ans\"atze systematically and determine whether the total-derivative phenomenon or the inversion symmetry persists.

\medskip
\mysec{Discussion}In this paper we have presented a ghost free and polynomially bounded
theory of scalars that is finite at all loop orders.  Our findings leave many
unanswered questions.  First and foremost is how or
whether this mechanism might be made phenomenologically viable for the Higgs
scalar.  To this end, some glaring obstacles include the sign of the
low-energy quartic, $\lambda<0$, as well as the incorporation of gauge
interactions.

A second line of inquiry concerns the origin of the momentum-dependent
interaction of the scalar.  Here a logically tenable viewpoint is that of Lee-Wick
models: classify the unitarity properties of the heavy
excitations while remaining agnostic about their detailed dynamics.  Of course, a
fuller picture would entail a complete microscopic description of the infinite-spin
modes that generate the scalar quartic.  But it is worth noting that a more elaborate
backstory for these states need not include self-interactions.
In principle, these infinite-spin modes could couple solely to the scalar in such a way that their only effect is the
momentum-dependent scalar interaction itself.  In this case they would resemble the massive
unstable particles in the standard model, which appear only as intermediate
resonances in the scattering of stable particles.

\medskip
\noindent {\it Acknowledgments}: We are grateful to Grant Remmen, Donal O'Connell, Yael Shadmi, and Ofri Telem for comments on the draft.  F.C. and C.C. are supported by the Department
of Energy (Grant No. DE-SC0011632), the Walter Burke Institute for Theoretical
Physics, and the Leinweber Forum for Theoretical Physics.


\medskip
\mysec{Appendix A: Finiteness Proof}We follow the method of Ref.~\cite{Weinberg1960} to rigorously establish ultraviolet finiteness for the scalar theories presented in this paper.  As is well-known, it is sufficient to show that the superficial degree of divergence is negative for any connected subdiagram or subintegration, since disconnected subdiagrams factorize into connected components.
  
\begin{figure}[h]
  \includegraphics[width=0.85\columnwidth]{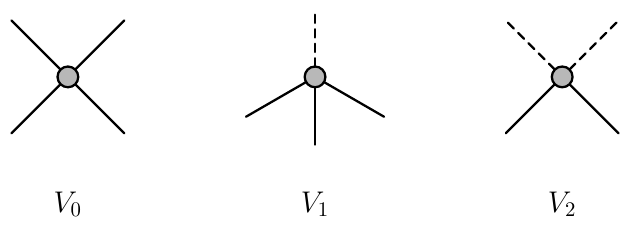}
  \caption{Interaction vertices classified by the number of soft (dashed) or hard (solid) legs attached. }
  \label{fig:vertices}
  \end{figure}

To begin, we stratify the quartic vertices according to their number of soft external legs, as depicted in \Fig{fig:vertices}.  The total number of vertices of the subdiagram is $V=V_0+V_1+V_2$. We also define $S=a_0V_0+a_1V_1+a_2V_2$ to be the degree of suppression from the form-factor interactions, relative to $\phi^4$ theory. Here the coefficients of each suppression term depend on the particular theory.

By definition, the single and double soft leg vertices appear on the boundary of a given hard subdiagram. Thus, the number of external legs satisfies $E=2V_2+V_1$, so it is bounded by $E\geq 2V_2$ and $E\geq V_1$.
The superficial degree of divergence of a subdiagram with $I$ internal edges in $D$ spacetime dimensions becomes
\eq{
\Delta&=DL-2I-S \\
&=(D-4)L+4-E-a_0V_0-a_1V_1-a_2V_2 \\
&=(D-4)L+4+a_0(1-L)-E\left(1+\frac{a_0}{2}\right)
\\[-1mm]
&\quad +(a_0-a_1)V_1+(a_0-a_2)V_2 \\
&\leq (D-4)L+4+a_0(1-L)-\frac{E}{2}(a_2+2+2a_1-2a_0),
}{eq:sdd}
where we used the inequalities above and the standard identities for any graph $V=I-L+1$, and for any graph with purely quartic vertices, $E=4V-2I$.  For the special case of $D=4$, \Eq{eq:sdd} reduces to
\eq{
\Delta\leq 4-(a_2+2+2a_1-2a_0),
}{eq:sdd-four-dim}
where we have assumed that the diagram has $E \geq 2$ external legs and $L\geq 1$ loops.

It is straightforward to verify that \Eq{eq:sdd} is valid in the scalar theories we have discussed. For the theory defined by \Eq{eq:one-mass-A}, we find that $a_0=a_1=6$ and $a_2=4$, so $\Delta \leq -2$. 
Meanwhile, for the theory defined by \Eq{eq:two-mass-A-tau}, we have $a_0=a_1=8$ and $a_2=4$, so $
\Delta\leq -2$.

To establish ultraviolet finiteness, we must also analyze subintegration regions. These are defined by taking some subset or linear combination of loop momenta to be hard,  with all complementary loop momenta soft.   Effectively, this procedure induces some number of constraints $C$ on the loop momentum vectors that define the integration region.  If this reroutes the loop momenta so that a previously hard internal line becomes soft, then we prune this internal line by treating it as a pair of soft external legs and then apply the arguments from above or the main text.   After pruning, we are left with a constrained loop integral which is less divergent than a generic hard loop integral by $DC$.  For the case of $D=4$, this constraint reduces the degree of divergence, relative to $\phi^4$ theory, by $4C$.  Since $\phi^4$ theory has at most quadratic divergences and $C\geq 1$, we deduce that the resulting contributions are finite.

\medskip

\mysec{Appendix B: Unitarity Bounds}In this section we determine the constraints from unitarity on the model in
\Eq{eq:two-mass-A-final}.  The residue at each resonance is
\eq{
R_\eta(x)
&=
\lim_{s\to \eta m^2}(\eta m^2-s)A(s,t,-s-t),
\\
&=
-\lambda m^2 c^2
\left[c-\frac{\eta^2}{4}(3+x^2)\right]
\\
&\quad\times
\left[
\left(\frac{c}{\eta}-\eta\right)
\prod_{\alpha=1,c}
\left(\alpha-\frac{\eta}{2}(x-1)\right)
\right.
\\
&\qquad\qquad\left.
\times
\left(\alpha+\frac{\eta}{2}(x+1)\right)
\right]^{-1},
}{eq:two-mass-residue-def}
where $\cos\theta=x=1+2t/{\eta m^2}$ for
$\eta=1,c$.
The partial wave coefficients are
\eq{
a_{\eta,\ell}
&=
\frac{2\ell+1}{2}
\int_{-1}^{1}dx\,
R_\eta(x)P_\ell(x),
\\
&=
-\frac{
2\lambda c^2m^2(2\ell+1)
}{
\eta(c-\eta^2)(c-1)(\eta+c+1)
}
\left[1+(-1)^\ell\right]\\
&\quad\times
\bigg[
(c-\eta^2-\eta-1)
\frac{
Q_\ell\!\left(1+\frac{2}{\eta}\right)
}{
1+\frac{2}{\eta}
}\\
&\qquad\quad
+
(c^2+c\eta+\eta^2-c)
\frac{
Q_\ell\!\left(1+\frac{2c}{\eta}\right)
}{
1+\frac{2c}{\eta}
}
\bigg],
}{eq:two-mass-partial-wave-proj}
where unitarity requires $a_{\eta,\ell} \geq 0$.  For $\lambda>0$ this cannot be satisfied, while for $\lambda \leq 0$ this implies that $c\ge 3$.

\bibliographystyle{apsrev4-1}
\bibliography{Ultraviolet_Finite_Scalar_inspire}

\end{document}